%% file: CPT-arXiv-v2.tex
\documentclass[twocolumn,showpacs,amsmath,amssymb,aps,prl,superscriptaddress]{revtex4}
\usepackage{enumerate}
\input{babarsym}
\newcommand{\BABARPubYear}    {16}
\newcommand{\BABARPubNumber}  {002}

\newcommand{\SLACPubNumber} {16525}

\newcommand{\Bqz}{{\overline B}{}^0}
\newcommand{\Kqz}{{\overline K}{}^0}
\newcommand{\cq}{{\overline c}}
\newcommand{\Aq}{{\overline A}}
\newcommand{\Bq}{{\overline B}}
\newcommand{\be}{\begin{equation}}
\newcommand{\ee}{\end{equation}}
\renewcommand{\Re}{{\rm Re\,}}
\renewcommand{\Im}{{\rm Im\,}}
\newcommand{\Af}{A_f}
\newcommand{\Afq}{{\overline A}{}_f}
\newcommand{\re}{{\rm e}}
\newcommand{\ri}{{\rm i}}
\newcommand{\bea}{\begin{eqnarray}}
\newcommand{\eea}{\end{eqnarray}}
\newcommand{\nn}{\nonumber}

\begin{document}

\begin{widetext}

\begin{flushleft}
\preprint{\babar-PUB-\BABARPubYear/\BABARPubNumber} 
\preprint{SLAC-PUB-\SLACPubNumber} 
\babar-PUB-\BABARPubYear/\BABARPubNumber\\
SLAC-PUB-\SLACPubNumber\\[20mm]
\end{flushleft}

\title{\bf{\boldmath Tests of  $CPT$ symmetry in $B^0$-$\Bqz$ mixing and in $B^0 \to c \cq K^0$ decays}}
\input authors_feb2016_frozen.tex

\begin{abstract} {\noindent
Using the eight time dependences $\re^{-\Gamma t}(1+C_i\cos\Delta m t+S_i\sin\Delta m t)$ for the decays
$\Upsilon(4S)\to B^0\Bqz\to f_jf_k$, with the decay into a flavor-specific state $f_j=\ell^\pm X$ before 
or after the decay into a $CP$ eigenstate $f_k=c\cq K_{S,L}$, 
as measured by the \babar experiment, we determine the three $CPT$-sensitive parameters $\Re({\mathsf z})$ 
and $\Im({\mathsf z})$ in $B^0$-$\Bqz$ mixing and $|\Aq/A|$ in $B^0\to c\cq K^0$ decays.
We find 
$\Im ({\mathsf z}) = 0.010 \pm 0.030 \pm 0.013$,
$\Re({\mathsf z}) = -0.065\pm 0.028\pm 0.014$,
and $|\Aq/A|=0.999\pm 0.023\pm 0.017$, in agreement with $CPT$ symmetry.}
\end{abstract}

\maketitle
\end{widetext}

\section{Introduction}

The discovery of $CP$ violation in 1964  \cite{1964-CroninFitch}  motivated searches for $T$ and $CPT$
violation. 
Since $CPT = CP \times T$, violation of $CP$  means that $T$ or $CPT$ or both are also violated. For the $K^0$ system,
the two contributions were first determined  \cite{1970-Schubert} in 1970, by using the Bell-Steinberger
unitarity relation \cite{1966-BellSteinberger} for $CP$ violation in $K^0$-$\Kqz$ mixing: 
$T$ was violated with about 5$\sigma$ significance
and no $CPT$ violation was observed. Large $CP$ violation in the $B^0$ system was discovered in 2001 
\cite{2001-Babar,2001-Belle} in the interplay of $B^0$-$\Bqz$ mixing and $B^0 \to c \cq K^0$ decays, but an 
explicit demonstration of $T$ violation was given  only recently \cite{2012-Lees}. 
In the present analysis, we test $CPT$ symmetry quantitatively in $B^0$-$\Bqz$ mixing 
and in $B^0 \to c \cq K^0$ decays.\\

Transitions in the $B^0$-$\Bqz$ system are well described by the quantum-mechanical evolution of a two-state wave function
\be 
\Psi=\psi_1~|B^0\rangle +\psi_2~|\Bqz\rangle ~,
\ee
using the Schr\"odinger equation
\be
{\dot{\Psi}}=-\ri~{\cal H}~\Psi~,
\ee
where the Hamiltonian $\cal H$ is given by two constant Hermitian matrices, ${\cal H}_{ij}=m_{ij}+\ri\Gamma_{ij}/2$. In this evolution,
$CP$ violation is described by three parameters, $|q/p|$, $\Re({\mathsf z})$, and $\Im({\mathsf z})$, defined by
\bea 
|q/p| &=& 1-\frac{2\, \Im (m^*_{12}\Gamma_{12})}{4|m_{12}|^2+|\Gamma_{12}|^2}~,\nn\\
{\mathsf z} &=& \frac {(m_{11}-m_{22})-\ri\,(\Gamma_{11}-\Gamma_{22})/2} {\Delta m-\ri\,\Delta\Gamma/2} \, ,
\eea
where $\Delta m=m(B_H)-m(B_L)\approx 2\,|m_{12}|$ 
and $\Delta\Gamma=\Gamma(B_H)-\Gamma(B_L)\approx +2\,|\Gamma_{12}|$ or $-2\,|\Gamma_{12}|$  
are the mass and the width differences of the two mass eigenstates ($H$=heavy, $L$=light) of the Hamiltonian,
\bea 
B_H &=& (p\,\sqrt{1+{\mathsf z}}\,B^0-q\,\sqrt{1-{\mathsf z}}\,\Bqz)/\sqrt{2}~,\nn\\
B_L &=& (p\,\sqrt{1-{\mathsf z}}\,B^0+q\,\sqrt{1+{\mathsf z}}\,\Bqz)/\sqrt{2}\label{Eq-4}\,.
\eea
Note that we use the convention with $+q$ for the light and $-q$ for the heavy eigenstate. 
If $|q/p|\ne 1$, the evolution violates the discrete symmetries $CP$ and $T$. If ${\mathsf z}\ne 0$, it violates $CP$ and $CPT$. 
The normalizations of the two eigenstates, as given in Eq.~(\ref{Eq-4}), are precise in the lowest order of $r$ and ${\mathsf z}$,
where $r=|q/p|-1$. Throughout the following, we neglect contributions of orders $r^2$, ${\mathsf z}^2$, $r\, {\mathsf z}$, 
and higher.\\

The $T$-sensitive mixing parameter $|q/p|$ has been determined in several experiments, the present world average 
\cite{2015-PDG} being $|q/p|=1+(0.8\pm0.8)\times 10^{-3}$. The $CPT$-sensitive parameter $\Im ({\mathsf z})$ has 
been determined by analyzing the time dependence of dilepton events in the decay 
$\Upsilon(4S)\to B^0\Bqz \to (\ell^+\nu X)~(\ell^-{\overline \nu} X)$; the \babar result \cite{2006-BABAR} is
$\Im ({\mathsf z}) = (-13.9\pm 7.3 \pm 3.2)\times 10^{-3}$. 
Since $\Delta\Gamma$ is very small, dilepton events are only sensitive to the product $\Re({\mathsf z})\Delta\Gamma$.
Therefore, $\Re({\mathsf z})$ has so far only been determined by analyzing the time dependence of the decays 
$\Upsilon(4S)\to B^0\Bqz$ with one $B$ meson decaying into $\ell\nu X$ and the other one into $c\cq K$. With 
$88\times 10^6 B\Bq$ events, \babar measured $\Re({\mathsf z})=(19\pm 48 \pm 47)\times 10^{-3}$ in 2004 \cite{2004-BABAR}, 
while Belle used $535\times 10^6 B\Bq$ events to
measure $\Re({\mathsf z}) =(19\pm 37 \pm 33)\times 10^{-3}$ in 2012 \cite{2012-Belle}.\\

In our present analysis, we use the final data set of the \babar experiment \cite{2002-BABAR,2013-BABAR-1} with
$470\times 10^6~B\overline B$ events for a new determination of $\Re({\mathsf z})$ and $\Im({\mathsf z})$. 
As in Refs.~\cite{2004-BABAR, 2012-Belle}, this is based on $c \cq K$ decays with amplitudes $A$ for 
$B^0\to c\cq K^0$ and $\Aq$ for $\Bqz\to c\cq \Kqz$, using the following two  assumptions:
\begin{enumerate}[(1)]
\item{$c\cq K$ decays obey the $\Delta S=\Delta B$ rule, i.\,e., $B^0$ states do not decay into $c\cq \Kqz$,
           and $\Bqz$ states do not decay into $c\cq K^0$;}
\item{$CP$ violation in $K^0$-$\Kqz$ mixing is negligible, i.\,e.\,$K^0_S=(K^0+\Kqz)/\sqrt{2}$, $K^0_L=(K^0-\Kqz)/\sqrt{2}$.}
\end{enumerate}

The $CPT$-sensitive parameters are determined from the measured time dependences of the four decay rates 
$B^0, \Bqz\to c \cq K^0_S, K^0_L$. In $\Upsilon(4S)$ decays, $B^0$ and $\Bqz$ mesons are produced in the entangled state 
$(B^0\Bqz-\Bqz B^0)/\sqrt{2}$. When the first meson decays into
$f=f_1$ at time $t_1$, the state collapses into the two states $f_1$ and $B_2$. The later decay $B_2\to f_2$ at 
time $t_2$ depends on the state $B_2$ and, because of $B^0$-$\Bqz$ mixing, on the decay-time difference 
\be
t=t_2-t_1\ge 0\, . \label{Eq-tDef}
\ee
Note that $t$ is the only relevant time here, it is the evolution time 
of the single-meson state $B_2$ in its rest frame.

The present analysis does not start from raw data but uses intermediate results from Ref.~\cite{2012-Lees} where, 
as mentioned above, we used our final data set for the demonstration of large $T$ violation. This was shown in 
four time-dependent transition-rate differences 
\be
R(B_j\to B_i) - R(B_i\to B_j)\, \label{Eq-TV}
\ee
where $B_i=B^0$ or $\Bqz$, and $B_j = B_+$ or $B_-$.
The two states $B_i$ were defined by flavor-specific decays \cite{DefFlavSpec} denoted as 
$B^0\to \ell^+ X$, $\Bqz\to \ell^- X$. The state $B_+$ was defined as the remaining state $B_2$ after a $c\cq K^0_S$
decay, and $B_-$ as $B_2$ after a $ c\cq K^0_L$ decay. In order to use the two states for testing $T$ symmetry 
in Eq.~(\ref{Eq-TV}), they must be orthogonal; $\langle B_+ | B_-\rangle =0$, which requires the additional assumption 
\begin{enumerate}[(3)]
\item $|\Aq/A| = 1$\, .
\end{enumerate}
In the same 2012 analysis, we demonstrated that $CPT$ symmetry is unbroken within uncertainties by measuring the four rate differences
\be 
R(B_j\to B_i) - R({\overline B}{}_i\to B_j)\, .\label{Eq-CPTT}
\ee
For both measurements in Eqs.~(\ref{Eq-TV}) and (\ref{Eq-CPTT}), expressions 
\be
R_i(t)=N_i~\re^{-\Gamma t}~(1+C_i\cos\Delta m t+S_i\sin\Delta mt),\label{Eq-RateDef} 
\ee
$i=1\ldots 8$, were fitted to the four time-dependent rates 
where the $\ell X$ decay precedes the
$c\cq K$ decay, and to the four rates where the order of the decays is inverted. The rate ansatz in Eq.~(\ref{Eq-RateDef}) requires
$\Delta\Gamma =0$. The time $t\ge 0$ in these expressions is the time between the first and the second decay 
of the entangled $B^0\Bqz$ pair as defined in Eq.~(\ref{Eq-tDef}). 
In our 2012 analysis, we named it $\Delta \tau$, equal to $t_{c\cq K}-t_{\ell X}$
if the $\ell X$ decay occurred first, and equal to $t_{\ell X}-t_{c\cq K}$ with $c\cq K$ as first decay. After the fits,
the $T$-violating and $CPT$-testing rate differences were evaluated from the obtained $S_i$ and $C_i$ results. 
The $CPT$ test showed no $CPT$ violation, i.\,e., it was compatible with ${\mathsf z}=0$, but no results for $\Re({\mathsf z})$ 
and $\Im({\mathsf z})$ were given in 2012.\\

Our present analysis uses the eight measured time dependences in
the 2012 analysis, i.\,e.\ the 16 results $C_i$ and $S_i$, for determining ${\mathsf z}$. This is possible without assumption (3)
since we do not need to use the concept of states $B_+$ and $B_-$. We are therefore able to 
determine the decay parameter $|\Aq/A|$ in addition to the mixing parameters  $\Re({\mathsf z})$ and $\Im({\mathsf z})$.
As in 2012, we use $\Delta\Gamma = 0$, but we show at the end of this analysis that the final results are independent of
this constraint.
Accepting assumptions (1) and (2), and in addition 
\begin{enumerate}[(4)]
\item{the amplitudes $A$ and $\Aq$ have a single weak phase,}
\end{enumerate}
only two more parameters $|\Aq/A|$ and $\Im(q\Aq/pA)$ are required
in addition to $|q/p|$ and ${\mathsf z}$ for a full description of $CP$ violation in time-dependent  $B^0\to c\cq K^0$ decays.
In this framework, $T$ symmetry requires $\Im(q\Aq/pA)=0$ \cite{1965-EnzLewis}, and $CPT$ symmetry requires $|\Aq/A|=1$ 
\cite{1957-LeeOehmeYang}.\\ 

\section{B-Meson Decay Rates}

The time-dependent rates of the decays $B^0,\Bqz\to c\cq K$ 
are sensitive to both symmetries $CPT$ and $T$ in $B^0$-$\Bqz$ mixing and in $B^0$ decays.
For decays into final states $f$ with amplitudes $A_f=A(B^0\to f)$ and $\Afq=A(\Bqz\to f)$, 
using $\lambda_f=q\Afq/(p\Af)$ and approximating $\sqrt{1-{\mathsf z}^2}=1$, the rates are given by
\begin{widetext}
\bea 
R(B^0\to f) &=& \frac{|A_f|^2\,\re^{-\Gamma t}}{4}\left|( 1-{\mathsf z}+\lambda_f)\,\re^{\ri\Delta m t}\,\re^{\Delta\Gamma t/4}
                                      +(1+{\mathsf z}-\lambda_f)\,\re^{-\Delta\Gamma t/4}\right|^2,\nn\\
R(\Bqz\to f) &=&  \frac{|\Afq|^2\,\re^{-\Gamma t}}{4}\left|( 1+{\mathsf z}+1/\lambda_f)\,\re^{\ri\Delta m t}\,\re^{\Delta\Gamma t/4}
                                      +(1-{\mathsf z}-1/\lambda_f)\,\re^{-\Delta\Gamma t/4}\right|^2.\label{Eq-5}
\eea
\end{widetext}
For the $CP$ eigenstates $c\cq K^0_L~(CP=+1)$ and  $c\cq K^0_S~(CP=-1)$ with $A_{S(L)}=A[B^0\to c\cq K^0_{S(L)}]$ and
$\Aq_{S(L)}=A[\Bqz\to c\cq K^0_{S(L)}]$, assumptions (1) and (2) give $A_S=A_L=A/\sqrt{2}$ and $\Aq_S=-\Aq_L=\Aq/\sqrt{2}$.  
In the following, we only need to use $\lambda_S=-\lambda_L=\lambda$.
Setting $\Delta\Gamma=0$ and keeping only first-order terms in the small quantities $|\lambda|-1$, ${\mathsf z}$, and $r=|q/p|-1$, this leads to
rate expressions as given in Eq.~(\ref{Eq-RateDef}) with coefficients
\begin{widetext}
\bea 
S_1=S(\ell^-X,c\cq K_L) &=& \frac{2\,\Im(\lambda)}{1+|\lambda|^2}-\Re({\mathsf z})\Re(\lambda)\Im(\lambda)+\Im({\mathsf z})[\Re(\lambda)]^2~,\nn\\
C_1&=& +\frac{1-|\lambda|^2}{2}- \Re(\lambda)\,\Re({\mathsf z})-\Im(\lambda)\,\Im ({\mathsf z})\,,\nn\\
S_2=S(\ell^+X,c\cq K_L) &=& -\frac{2\,\Im(\lambda)}{1+|\lambda|^2}-\Re({\mathsf z})\Re(\lambda)\Im(\lambda)-\Im({\mathsf z})[\Re(\lambda)]^2~,\nn\\
C_2 &=&- \frac{1-|\lambda|^2}{2} +\Re(\lambda)\,\Re({\mathsf z})-\Im(\lambda)\,\Im ({\mathsf z})\,,\nn\\ 
S_3=S(\ell^-X,c\cq K_S) &=& -\frac{2\,\Im(\lambda)}{1+|\lambda|^2}-\Re({\mathsf z})\Re(\lambda)\Im(\lambda)+\Im({\mathsf z})[\Re(\lambda)]^2~,\nn\\
C_3 &=& +\frac{1-|\lambda|^2}{2}+\Re(\lambda)\,\Re({\mathsf z})+\Im(\lambda)\,\Im ({\mathsf z})\,,\nn\\
S_4=S(\ell^+X,c\cq K_S) &=& \frac{2\,\Im(\lambda)}{1+|\lambda|^2}-\Re({\mathsf z})\Re(\lambda)\Im(\lambda)-\Im({\mathsf z})[\Re(\lambda)]^2~,\nn\\
C_4 &=& -\frac{1-|\lambda|^2}{2}-\Re(\lambda)\,\Re({\mathsf z})+\Im(\lambda)\,\Im ({\mathsf z})\,.\label{Eq-8}
\eea
\end{widetext}
The four other rates  $R_5(t) \cdots R_8(t)$ with $c\cq K$ as the first decay and $t_{\ell X} - t_{c\cq K}=t$ follow from 
the same two-decay-time expression \cite{twodecaytime1,twodecaytime2} as the rates $R_1\dots R_4$ with $t_{c\cq K}-t_{\ell X}=t$.
Therefore, the rates $R_5(c\cq K_L,\ell^-X)$, $R_6(c\cq K_L,\ell^+X)$, $R_7(c\cq K_S,\ell^-X)$, and $R_8(c\cq K_S,\ell^+X)$ are given by
Eq.~(\ref{Eq-RateDef}) with the coefficients
\be
S_i = -S_{i-4}~,~~C_i=+C_{i-4}~~{\rm for~}i=5,6,7,~{\rm and}~8\,.\label{Eq-9}
\ee
The $S_i$ and $C_i$ results from our 2012 analysis, including uncertainties and correlation matrices, have been published as
Supplemental Material \cite{2012-Lees-appendix} in Tables II, III, and IV. 
For completeness, we include in Table \ref{Tab-1-inputs} the results and the uncertainties.
\begin{table}[h!]
\setlength{\belowcaptionskip}{2mm}
\caption{Input values from the Supplemental Material \cite{2012-Lees-appendix} of 
Ref.~\cite{2012-Lees}. The second column gives the two decays with their sequence in decay time.}
\label{Tab-1-inputs}
\begin{center}
\begin{tabular}{c c c r r r r r r r r}
\hline\hline
$i$ & & decay pairs & & $S_i$ & $\sigma_{\rm stat}$ & $\sigma_{\rm sys}$ & & $C_i$ & $\sigma_{\rm stat}$ & $\sigma_{\rm sys}$ \\
\hline\hline
1 & & $\ell^- X,c\cq K_L$ & & 0.51   & 0.17 & 0.11 & & $-0.01$ & 0.13 & 0.08\\
2 & & $\ell^+ X,c\cq K_L$ & & $-0.69$ & 0.11 & 0.04  & & $-0.02$ & 0.11 & 0.08\\
3 & & $\ell^- X,c\cq K_S$ & & $-0.76$ & 0.06 & 0.04  & & 0.08 & 0.06 & 0.06\\
4 & & $\ell^+ X,c\cq K_S$ & & 0.55 & 0.09 & 0.06  & & 0.01 & 0.07 & 0.05\\
5 & & $c\cq K_L,\ell^- X$ & & $-0.83$ & 0.11 & 0.06 & & 0.11 & 0.12 & 0.08\\
6 & & $c\cq K_L,\ell^+ X$ & & 0.70 & 0.19 & 0.12 & & 0.16 & 0.13 & 0.06\\
7 & & $c\cq K_S,\ell^- X$ & & 0.67& 0.10 & 0.08 & & 0.03 & 0.07 & 0.04\\
8 & & $c\cq K_S,\ell^+ X$ & & $-0.66$ & 0.06 & 0.04 & & $-0.05$ & 0.06 & 0.03\\
\hline\hline\end{tabular}
\end{center}
\end{table}

\section{Fit Results}

The relations between the 16 observables $y_i = S_1 \cdots C_8$ in Eqs.~(\ref{Eq-8}) and (\ref{Eq-9}) and the four parameters 
$p_1=(1-|\lambda|^2)/2$, $p_2=2\,\Im(\lambda)/(1+|\lambda|^2)$, $p_3=\Im ({\mathsf z})$, and
$p_4=\Re({\mathsf z})$ are approximately linear. Therefore, the four parameters can be determined in a two-step 
linear $\chi^2$ fit using matrix algebra.
The first-step fit determines $p_1$ and $p_2$ by fixing $\Re(\lambda)$ and $\Im(\lambda)$
in the products $\Re({\mathsf z})\Re(\lambda)$, $\Im({\mathsf z})\Im(\lambda)$, $\Im({\mathsf z})[\Re(\lambda)]^2$, and
$\Re({\mathsf z})\Re(\lambda)\Im(\lambda)$. After fixing these terms, the relation between the vectors $y$  and $p$  is strictly linear,
\be y = M_1~p,  \label{Eq-fitA}\ee
where $M_1$ uses $\Im(\lambda)=0.67$ and $\Re(\lambda)=-0.74$, motivated by the results of analyses assuming $CPT$ symmetry
\cite{2015-PDG}.
With this ansatz, $\chi^2$ is given by
\be \chi^2 = (M_1~p-\hat{y})^T~G~(M_1~p - \hat{y}), \ee
where $\hat{y}$ is the measured vector of observables, and the weight matrix $G$ is taken to be
\be G = [C_{\rm stat}(y)+C_{\rm sys}(y)]^{-1}~, \ee
where $C_{\rm stat}(y)$ and $C_{\rm sys}(y)$ are the statistical and systematic covariance matrices, respectively. 
The minimum of $\chi^2$ is reached for
\be \hat{p} = {\cal M}_1~\hat{y}~~{\rm with}~~ {\cal M}_1=(M_1^T~G~M_1)^{-1}~M_1^T~G~,\ee
and the uncertainties of $\hat{p}$ are given by the covariance matrices
\bea 
C_{\rm stat}(p) &=& {\cal M}_1~C_{\rm stat}(y)~{\cal M}_1^T~,\nn\\
C_{\rm sys}(p) &=& {\cal M}_1~C_{\rm sys}(y)~{\cal M}_1^T~, 
\eea
with the property
\be C_{\rm stat}(p) +C_{\rm sys}(p) =(M_1^T~G~M_1)^{-1}~.\label{Eq-fitE}\ee
This first-step fit yields
\bea
p_1 &=& 0.001 \pm 0.023 \pm 0.017\, ,\nn\\
p_2 &=& 0.689 \pm 0.030 \pm 0.015\, .\label{Eq-finalCS}
\eea
This leads to
\bea
|\lambda| &=& 1-p_1 = 0.999 \pm 0.023 \pm 0.017\,,\nn\\
\Im(\lambda) &=& (1-p_1)\,p_2  = 0.689\pm 0.034 \pm 0.019\, ,\nn\\
\Re(\lambda) &=& -(1-p_1)\,\sqrt{1-p_2^2} \nn\\
 &=& -0.723 \pm 0.043 \pm 0.028\, ,\label{Eq-finalLambda}
\eea
where the negative sign of $\Re(\lambda) $ is motivated by four measurements 
\cite{2005-Aubert,2005-Belle,2007-BABAR-1,2007-BABAR-2}. The results of all four favor
$\cos 2\beta >0$, and in Ref.~\cite{2007-BABAR-2} $\cos 2\beta <0$ is excluded with 4.5 $\sigma$ significance.\\

In the second step,
we fix the two $\lambda$ values according to the $p_1$ and $p_2$ results of the first step, i.e. to the central values in
Eqs.~(\ref{Eq-finalLambda}). Equations (\ref{Eq-fitA}) to (\ref{Eq-fitE}) are then applied again, replacing $M_1$ with the new
relations matrix $M_2$. This gives 
the same results for $p_1$ and $p_2$ as in Eq.~(\ref{Eq-finalCS}), and
\bea
p_3=\Im ({\mathsf z})&=&0.010 \pm 0.030 \pm 0.013\,,\nn\\
p_4=\Re({\mathsf z})&=&-0.065\pm 0.028\pm 0.014 \, ,\label{Eq-FinalZ}
\eea
with a $\chi^2$ value of 6.9 for 12 degrees of freedom.\\

The $\Re({\mathsf z})$ result deviates from $0$ by $2.1\, \sigma$.
The result for $|\lambda|$ can be easily  converted into $|\Aq/A|$ by using the world average of measurements for
$|q/p|$. With $|q/p|=1.0008\pm0.0008$  \cite{2015-PDG}, we obtain 
\be
|\Aq/A|= 0.999\pm 0.023\pm 0.017~,
\ee
in agreement with $CPT$ symmetry.
Using the matrix algebra in Eqs.~(\ref{Eq-fitA}) to (\ref{Eq-fitE}) allows us to determine the separate statistical and systematic covariance 
matrices of the final results, in agreement with the condition $C_{\rm stat}(p)+C_{\rm sys}(p)=(M^T~G~M)^{-1}$, where $M$ relates $y$ and
$p$ after convergence of the fit.
The statistical correlation coefficients are $\rho[|\Aq/A|, \Im({\mathsf z})]=0.03$, $\rho[|\Aq/A|, \Re({\mathsf z})]=0.44$, and
$\rho[\Re({\mathsf z}), \Im({\mathsf z})]=0.03$. The systematic correlation coefficients are $\rho[|\Aq/A|, \Im({\mathsf z})]=0.03$, 
$\rho[|\Aq/A|, \Re({\mathsf z})]=0.48$, and $\rho[\Re({\mathsf z}), \Im({\mathsf z})]=-0.15$.\\

\section{\boldmath Estimating the Influence of $\Delta\Gamma$}

Using an accept/reject algorithm,
we have performed two ``toy simulations", each with $\sim 2\times 10^6$ events, i.e. $t$ values sampled 
from the distributions
\be 
\re^{-\Gamma t}[1+\Re(\lambda)\sinh(\Delta\Gamma\,t/2)+\Im(\lambda)\sin(\Delta m\,t)]~,
\ee
with $\Delta\Gamma =0$ for one simulation and $\Delta\Gamma =0.01\Gamma$  for the other one, corresponding to one 
standard deviation from the present world average \cite{2015-PDG}. For both simulations we use $\Im(\lambda)=0.67$ and $\Re(\lambda)=-0.74$
and sample $t$ values between $0$  and  $+5/\Gamma$.
We then fit the two samples, binned in intervals of $\Delta t=0.25/\Gamma$, to the expressions 
\be 
N \re^{-\Gamma t}[1+C \cos(\Delta m\, t)+S \sin(\Delta m\, t)]~,
\ee
with three free parameters $N$, $C$ and $S$.
The fit results 
agree between the two simulations within $0.002$ for $C$ and $0.008$ for $S$. 
We, therefore, conclude that omission of the sinh term
in Ref.~\cite{2012-Lees} has a negligible influence on the three final results of this analysis.\\

\section{Conclusion}

Using $470\times 10^6~B\overline B$ events from \babar, we determine
\bea
\Im ({\mathsf z}) &=& 0.010 \pm 0.030 \pm 0.013\,,\nn\\
\Re({\mathsf z}) &=& -0.065\pm 0.028\pm 0.014 \, ,\nn\\
|\Aq/A| &=& 0.999\pm 0.023\pm 0.017\,,\nn
\eea
where the first uncertainties are statistical and the second uncertainties are systematic.
All three results are compatible with $CPT$ symmetry in $B^0$-$\Bqz$ mixing and in $B\to c\cq K$ decays. The
uncertainties on $\Re({\mathsf z})$ are comparable with those obtained by Belle in 2012 
 \cite{2012-Belle} with $535\times 10^6~B{\overline B}$ events, $\Re({\mathsf z}) = -0.019 \pm 0.037 \pm 0.033$.
The uncertainties on $\Im ({\mathsf z})$ are considerably larger, as expected, than those obtained
by \babar in 2006 \cite{2006-BABAR} with dilepton decays from $232\times 10^6~{B\overline B}$ events, 
$\Im ({\mathsf z}) = -0.014\pm 0.007\pm 0.003$. The result of the present analysis for $\Re({\mathsf z}),\,
-0.065\pm 0.028\pm 0.014$,
supersedes the \babar result of 2004 \cite{2004-BABAR}.\\

\section{Acknowledgements}

%Short version for PRL/PRD-RC is used
We thank
H.-J.~Gerber (ETH Zurich) and T.~Ruf (CERN) for very useful discussions on $T$ and $CPT$ symmetry.
\input{acknow_PRL.tex}

\raggedright

\end{document}

%% file: authors_feb2016_frozen.tex
% NOTES
%
% 20-FEB-2016 Add footnote for Liang Sun                               J.W. Gary
% 21-DEC-2015 Add Bologna alternative address for Claudia Patrignani   J.W. Gary
%
\author{J.~P.~Lees}
\author{V.~Poireau}
\author{V.~Tisserand}
\affiliation{Laboratoire d'Annecy-le-Vieux de Physique des Particules (LAPP), Universit\'e de Savoie, CNRS/IN2P3,  F-74941 Annecy-Le-Vieux, France}
\author{E.~Grauges}
\affiliation{Universitat de Barcelona, Facultat de Fisica, Departament ECM, E-08028 Barcelona, Spain }
\author{A.~Palano}
\affiliation{INFN Sezione di Bari and Dipartimento di Fisica, Universit\`a di Bari, I-70126 Bari, Italy }
\author{G.~Eigen}
%\author{B.~Stugu}
\affiliation{University of Bergen, Institute of Physics, N-5007 Bergen, Norway }
\author{D.~N.~Brown}
%\author{L.~T.~Kerth}
\author{Yu.~G.~Kolomensky}
%\author{M.~J.~Lee}
%\author{G.~Lynch}
\affiliation{Lawrence Berkeley National Laboratory and University of California, Berkeley, California 94720, USA }
\author{H.~Koch}
\author{T.~Schroeder}
\affiliation{Ruhr Universit\"at Bochum, Institut f\"ur Experimentalphysik 1, D-44780 Bochum, Germany }
% ...
\collaboration{The \babar\ Collaboration}
\noaffiliation

%% file: acknow_PRL.tex
We are grateful for the excellent luminosity and machine conditions
provided by our \pep2\ colleagues, 
and for the substantial dedicated effort from
the computing organizations that support \babar.
The collaborating institutions wish to thank 
SLAC for its support and kind hospitality. 
This work is supported by
DOE
and NSF (USA),
NSERC (Canada),
CEA and
CNRS-IN2P3
(France),
BMBF and DFG
(Germany),
INFN (Italy),
FOM (The Netherlands),
NFR (Norway),
MES (Russia),
MINECO (Spain),
STFC (United Kingdom),
BSF (USA-Israel). 
Individuals have received support from the
Marie Curie EIF (European Union)
and the A.~P.~Sloan Foundation (USA).
% NOTES:
% add "and the Binational Science Foundation (U.S.-Israel)"  07-Oct-2013 Bill Gary (Abi Soffer request)